\documentclass[twocolumn,preprintnumbers,amsmath,amssymb]{revtex4}
\usepackage{graphicx}
\usepackage{bm}
\usepackage{dcolumn}
\usepackage{epsfig}

\newcommand{\A}{A$_2$Ti$_2$O$_7$ (A=Eu, Gd, Tb, Dy, Ho, Er, Yb)}
\newcommand{\Aa}{A$_2$Ti$_2$O$_7$ (A=Eu, Gd, Tb, Dy, Yb)}

\newcommand{\Ae}{A$_2$Ti$_2$O$_7\,$}
\newcommand{\Eu}{Eu$_2$Ti$_2$O$_7\,$}
\newcommand{\Dy}{Dy$_2$Ti$_2$O$_7\,$}
\newcommand{\Gd}{Gd$_2$Ti$_2$O$_7\,$}
\newcommand{\Tb}{Tb$_2$Ti$_2$O$_7\,$}
\newcommand{\Ho}{Ho$_2$Ti$_2$O$_7\,$}
\newcommand{\Er}{Er$_2$Ti$_2$O$_7\,$}
\newcommand{\Yb}{Yb$_2$Ti$_2$O$_7\,$}

\newcommand{\non}{\nonumber}
\begin{document}

\title{First-principles insights into f magnetism, 
       a case study on some magnetic pyrochlores}

\author{Najmeh Deilynazar}

\author{Elham Khorasani}

\author{Mojtaba Alaei}
\email{m.alaei@cc.iut.ac.ir}

\author{S. Javad Hashemifar}

\affiliation{Department of Physics, Isfahan University of
Technology, Isfahan 84156-83111, IRAN.}

\date{\today}

\begin{abstract}

First-principles calculations are performed to investigate f magnetism 
in \A~magnetic pyrochlore oxides.
The Hubbard U parameter and the relativistic spin orbit correction
is applied for more accurate description of the electronic structure of the systems.
It is argued that the main obstacle for first-principles study of these systems
is the multi-minima solutions of their electronic configuration.
Among the studied pyrochlores, \Gd shows the least multi-minima problem.
The crystal electric field theory is applied for phenomenological
comparison of the calculated spin and orbital moments with the experimental data.

\end{abstract}
\maketitle

\section{Introduction}

\begin{figure}
\centering
\includegraphics[width=0.45\columnwidth]{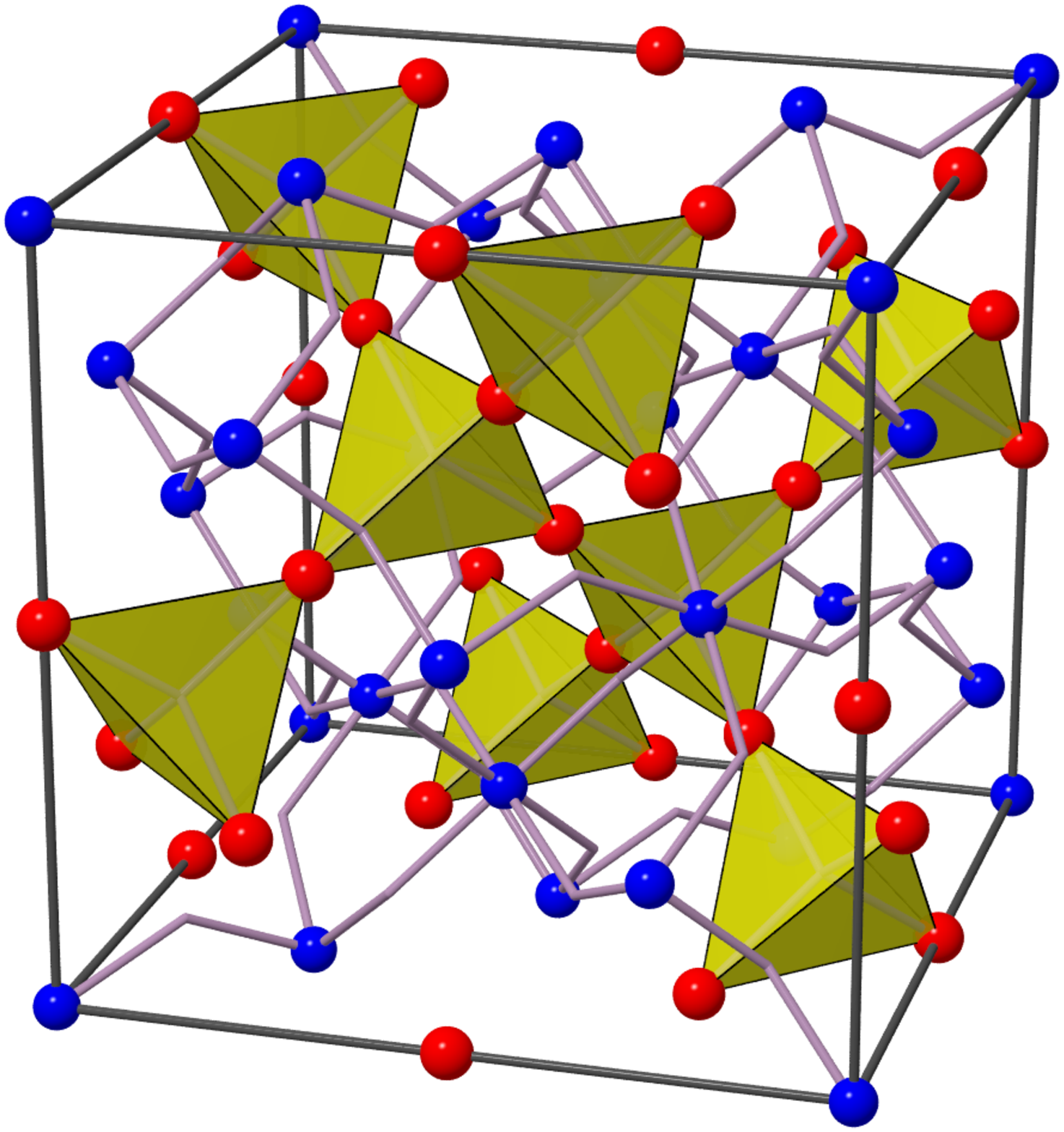}
\includegraphics[width=0.45\columnwidth]{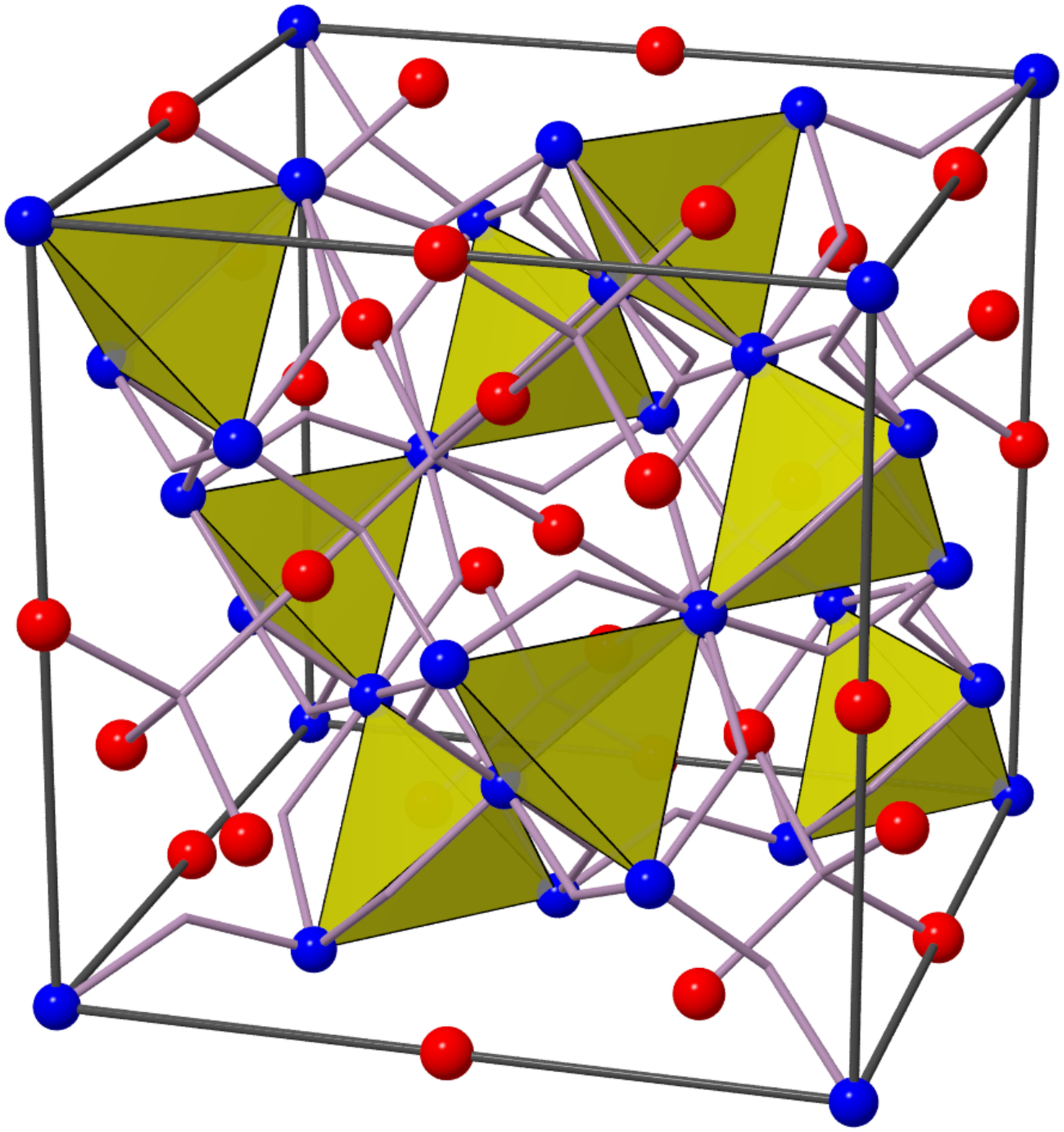}
\caption{
 The structure of pyr-$\mathrm{A}_2\mathrm{Ti}_2\mathrm{O}_7$.
 Left: positions of magnetic A$^{3+}$ ions (red color) and  Right: positions
 of Ti$^{4+}$ ions (blue color) on the the cubic pyrochlore lattice. The 
 O$^{-2}$ ions (not shown) are located at the bent (in gray) where bonds merge.}
\label{struct}
\end{figure}

Magnetic pyrochlore oxides~\cite{gardner2010magnetic} with chemical formula
A$_2$B$_2$O$_7$ have rich physics and exotic magnetic
properties (such as magnetricity\cite{Bramwell2009}) caused by geometrical frustration. 
In these materials A and B are
usually trivalent rare-earth and travalent transition metal ions, respectively,
which form a network of corner sharing tetrahedral on 
the fcc Bravais lattice (Fig.~\ref{struct}).
This geometrical feature is known to be the origin of 
the magnetic frustration of the system.
Among these materials, \Ae~(A=Eu, Gd, Tb, Dy, Ho, Er, Yb) 
are the most studied compounds \cite{petrenko2011titanium},
and exhibit various magnetic phenomena including spin ice behavior 
in \Ho~ and \Dy~\cite{gardner2010magnetic}, 
spin liquid behaviour in \Tb~\cite{gardner2001neutron,gardner1999cooperative} 
and \Yb~\cite{hodges2002first}, 
and order by disorder phenomena in 
\Er~\cite{gardner2010magnetic,bramwell2000bulk,poole2007magnetic}.
\Gd, as a controversial case, is assumed to be a classical Heisenberg 
antiferromagnet~\cite{raju1999transition}.

Theoretical investigations on magnetic pyrochlores are mainly
based on model Hamiltonians, involving Heisenberg exchange, 
dipole-dipole interactions, and single ion anisotropy, 
which are parametrized by using experimental data \cite{gardner2010magnetic}.
The lack of modern ab initio calculations on these materials 
prevents microscopic understanding of these magnetic model Hamiltonians.
To our knowledge, ab initio calculations on magnetic pyrochlore oxides
are limited to nonmagnetic properties \cite{xiao2008first,xiao2013electronic} 
or those compounds
with no active f electrons \cite{nemoshkalenko2001electronic}. 
It is due to the fact that the partially filled 4f orbital of the rare earth ions 
involves strongly correlated and localized electrons 
which cause serious difficulties in finding the true ground state of the system
within density functional theory (DFT) computations~\cite{localminima}.
It is well understood that conventional local functionals
give rise to wrong ground states for these strongly correlated systems.
The usual solution for correct description of the coulomb interaction 
between 4f electrons is applying orbital dependent approaches including  
hybrid functionals~\cite{becke1993new} and the Hubbard based DFT+U technique.
An important challenge in first-principles calculation of 
the systems with 4f electrons within orbital dependent functionals 
is sensitivity of the final results to the initial electronic
configuration of the 4f shell.
Moreover, close energy local minima are serious obstacles to
achieve convergency in first-principles calculation of the 4f electron systems.

Our specific aim in this work is employing DFT+U approach for 
first-principles calculation and investigation of f magnetism
in \A~ compounds.
After brief explanation of our computational method,
the obtained results are presented in section III.
In section IV, our conclusions are presented.

\section{Computational method}

Our electronic structure calculations are performed by using 
Fleur \cite{FLEURgroup:Misc} computer code, which solves 
Kohn-Sham single particle equations by using the full-potential 
linearized augmented plane wave (FP-LAPW) method ~\cite{wimmer1981full}.
The muffin-tin radii of A, Ti, and O atoms were set to 2.75, 2.0, 
and 1.35 bohr, respectively. 
The Perdew-Burke-Ernzerhof (PBE) formulation of 
the generalized gradient functional (GGA) is used in this work~\cite{pbe}.
We used the GGA+U approximation for better description of 
the Coulomb interaction between f electrons. 
The adapted values of U for \Ae~are given in table~\ref{gap},
while the value of on-site Hund's exchange, J$_H$, is set to 1 eV.
The spin-orbit coupling (SOC) is very important in 
the systems with heavy 4f elements, 
hence we considered this relativistic interaction in our calculations.

\begin{table}
\newcommand{\ra}{\cite{pandit1992electrical}}
\newcommand{\rb}{\cite{parida2011visible}}
\newcommand{\rc}{\cite{Pandit199152}}
\newcommand{\rd}{\cite{bramwell2001spin}}
\caption{\label{gap}
 Hubbard U parameters used in this work (taken from Ref.~\cite{Sannathesis}),
 calculated band gap (this work),
 lattice constant ($a$), and structural internal parameter of
 oxygen ($x$) \cite{lian2003radiation} 
 in the studied titanate pyrochlore oxides.
 The available experimental values of the band gaps
 are given in the parenthesis.} 
\begin{ruledtabular}
\begin{tabular}{cclcc}
    & U (eV) & gap (eV)  & $a~(\AA)$ &  $x$  \\               
\hline
\Eu & 10.14 &   2.1 (2.5 \ra) & 10.194 & 0.327  \\
\Gd & 11.48 &   3.0 (3.2 \rb) & 10.186 & 0.326  \\
\Tb &  5.00 &   2.4           & 10.159 & 0.328  \\
\Dy &  5.68 &   1.9 (2.4 \rc) & 10.124 & 0.328  \\
\Ho &  6.82 &   ---~ (3.2 \rd)& 10.104 & 0.329  \\
\Er &  6.80 &   3.0           & 10.071 & 0.328  \\
\Yb &  6.00 &   0.6           & 10.325 & 0.331  \\
\end{tabular}
\end{ruledtabular}
\end{table}

A $4\times 4\times 4$ Monkhorst-Pack k mesh 
was used for Brillouin zone integration. 
The lanthanide $4f$ electrons were treated as valance states while
their $5s^{2} 5p^{6}$ electrons as well as the Ti $3s^2 3p^6$ electrons 
were considered as semi-core states.
The cut-off of wave function expansion in the interstitial region 
was set to 3.8~$(a.u)^{-1}$. 
We adapted experimental lattice parameters of \Ae,
given in Table \ref{gap}, for our calculations. 
The space group of A$_2$Ti$_2$O$_7$ is
Fd$\overline{3}$m, with the following Wyckoff positions
for A and Ti atoms, receptively;
16d $(1/2,1/2,1/2)$, Ti: 16c (0,0,0).
There are two Wyckoff positions for oxygen; 
O at 48f $(x,1/8,1/8)$, and O$'$ at 8b $(3/8,3/8,3/8)$.
The primitive cell of \Ae, contains 22 atoms
and the internal parameter $x$ was taken from experiment (Table~\ref{gap}).

Unfortunately the calculations for \Er and \Ho~ were unstable and we could 
not achieve their converged electronic structure within GGA+U+SOC. 
Hence in the case of \Er, we report our 
previous converged results~\cite{Khorasani} within LDA+U+SOC
and muffin-tin radii 2.5, 2, 1.5 bohr for Er, Ti, and O, respectively.
While for Holmium Titanate, we could converge its electronic structure
within GGA+U to determine its band gap and spin moment.

\begin{figure}
\includegraphics*[scale=.7]{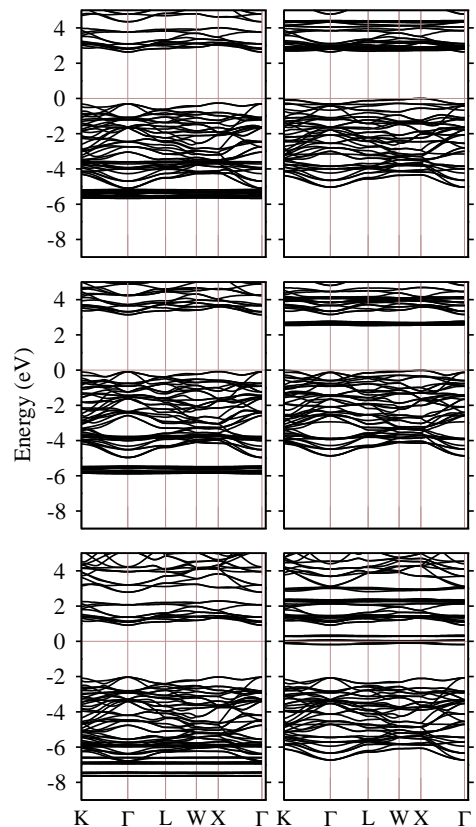}
\caption{
 Obtained three different band structures of \Tb~within GGA+U.
 Please see text for more details.
 The Fermi energies are set to zero.}
\label{picture2} 
\end{figure}

\section{Results}

As it was mentioned, the multi minima problem is a serious challenge
in first-principles investigation of 4f electron systems.
In order to see the existence of this problem, we performed some 
test calculations on \Tb~within GGA+U, by starting from 
three different initial states.
In these preliminary calculations, we omitted the SOC interaction,
while other parts of our study involves this relativistic correction.
At the end of calculations, we reached to three different electronic band 
structures, presented in Fig.~\ref{picture2}. 
The top band structure is obtained
by using the full symmetry (Fd$\overline{3}$m) of the system
during the computations, 
while the middle band structure is calculated without using any symmetry.
It is clearly visible that in the resulted band structure after
low symmetry calculations,
the narrow occupied and unoccupied $f$ bands are slightly separated from 
the valence and conduction bands.
The bottom band structure was calculated by using the full symmetry 
of the structure and the converged metallic electronic structure 
of the system within GGA as the starting point of GGA+U calculation.
In this way, the GGA+U calculation is converged to a 
metallic electronic structure.

\begin{figure}
\includegraphics*[scale=0.95]{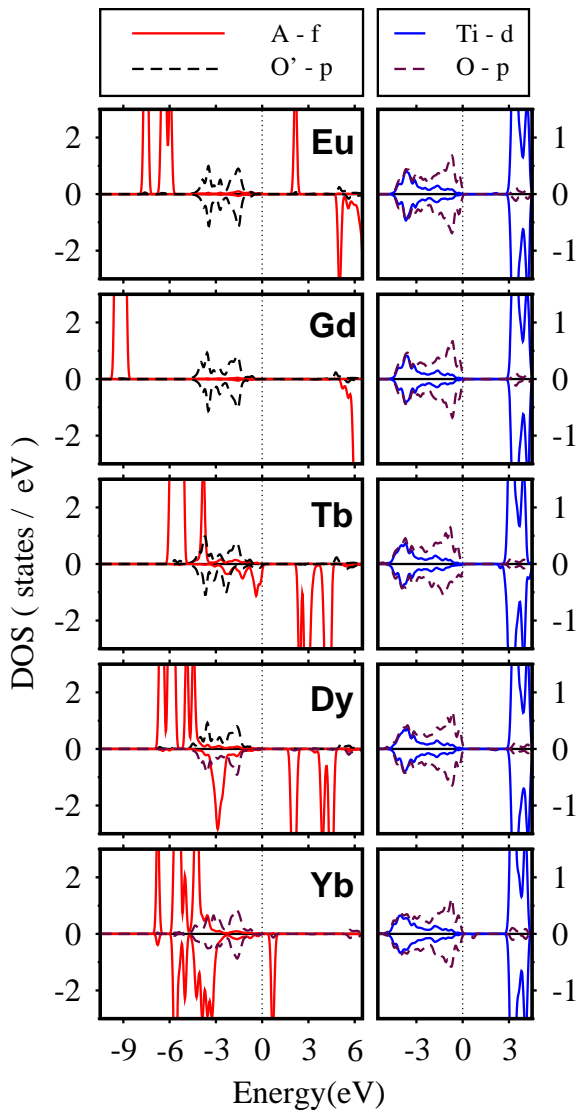}
\caption{\label{dos} 
 Calculated partial density of states (DOS) of the studied titanate 
 pyrochlore \Aa$\,$ by GGA+U+SOC. The Fermi energies are set to zero.} 
\end{figure}

As it was mentioned, GGA predicts a wrong metallic state for most of the 4f compounds, 
while GGA+U opens a gap between f states and make \Ae~insulators. 
The calculated band gap of these compounds within GGA+U+SOC
are given in Table~\ref{gap}, along with the available experimental data. 
The good agreement observed between computed and measured band gaps
confirms the selected U parameters for these materials.
The obtained orbital resolved Density of State (DOS) of 
the studied systems are shown in Fig.~\ref{dos}.
As it was mentioned in Computational method, 
there are two kinds of oxygen atoms in the systems.
The first kind of O atoms connect the neighboring Ti$^{4+}$ ions displayed in 
Fig. \ref{struct} and hence, a good hybridization happens between d-orbital of Ti and 
p-orbital of these O atoms (Fig. \ref{dos}). 
The second kind of O atoms (O$'$) are inside tetrahedrons of A$^{3+}$ ions,
hence their p orbital DOS are compared with the f orbital DOS of A atoms in Fig. \ref{dos}.
The hybridization between these orbitals is clearly week, which is likely due to 
the localized nature of f states.
However, in the case of Tb, Dy, and Yb based pyrochlores, the contribution of f states 
in the valence shell is high, showing the importance of valence treatment of f electrons 
for ab initio calculation of these systems.  
It is seen that among \Ae, \Gd~exhibits the lowest contribution of f states
in the valence shell, which is due to the half-filling of f orbital in Gd.
As a result of that, there is only one set of possible occupation numbers in \Gd
and this helps the ab initio calculations to easily converge to its global minima.

The calculated spin, orbital, and total magnetic moments of the studied \Ae~pyrochlores
obtained from collinear GGA+U+SOC calculations are given in Table \ref{magnetic}.
These moments are calculated inside Muffin-tin spheres. 
The spin quantization axis is set to the z direction. 
The experimental values of total moments are also reported for comparison. 
The calculated effective magnetic moment of \Eu
is about 4.9 ${\mu_{B}}$, which shows large deviation from 
the experimental value of 0.6$\mu_{B}$ \cite{dasgupta2007low}. 
In contrast, the calculated energy gap of this system is 
close to the experimental value (Table \ref{gap}).
In the case of \Gd~we got a magnetic moment of 7.0$\mu_{B}$,
in good agreement with the measured value of 7.7$\mu_{B}$ \cite{raju1999transition}). 
The $Gd^{3+}$ ion with a valence shell of $(4f^{7})$ 
has a spin of $S=7/2$ with no orbital magnetic moment, 
giving rise to an effective moment of $g\times\sqrt{S(S+1)}=7.9$~(g=2), 
in good agreement with the experimental moment.
The GGA band gap of \Gd~is found to be about 2.7 eV, in agreement 
with Xiao \textit{et al.} \cite{xiao2007theoretical}.
This compound is the only titanate pyrochlore with a band gap within GGA. 
This observation provides further evidence for the fact that
this system easily converges to its true ground state during ab initio calculations.
It should be admitted that the true ground state of this system, as well as other
magnetic pyrochlores, may be a non-collinear antiferromagnetic state.
Our preliminary calculations confirm the non-collinear ground state of \Gd,
however accurate determination of this ground state requires
comprehensive ab initio calculations which is out of the scope of this paper.
It should be noted that non-collinear calculations for other pyrochlores
are much more complicated.

In \Tb, GGA+U+SOC gives a total moment of about 7.2$\mu_{B}$,
which is larger than the measured value of 5.1$\mu_{B}$~\cite{gardner2001neutron}.
On the other hand, the calculated total moment of \Dy (8.2$\mu_{B}$) is lower 
than the experimental moment of 10$\mu_{B}$ \cite{gardner2010magnetic}.
In the case of \Yb, our ab initio calculation predicts 
a moment of 2.4 $\mu_{B}$ which is about twice the experimental value
of about 1.1$\mu_{B}$ measured by single-crystal neutron diffraction 
method below 0.2 K ~\cite{hodges2002first}. 
As it was mentioned in Computational method, we could not converge
our GGA+U+SOC calculations for \Er~and \Ho.
The total moment of Erbium Titanate within LDA+U+SOC was found to be
8.8 $\mu_{B}$, which is about 5.7 $\mu_{B}$ bigger than 
the experimental value \cite{poole2007magnetic}.
The spin moment of \Ho~within GGA+U was found to be about 3.8 $\mu_{B}$

\begin{table*}
\caption{
 Magnetic properties of titanate pyrochlore,
 L,S,J: orbital, spin, and total angular momentum,
 $\mu_{S}$, $\mu_{L}$ ($\mu_B$): Calculated spin and orbital magnetic moments,
 The values in the parenthesis are estimated from crystal field analysis.
 $\mu_{tot}$ ($\mu_B$): Calculated total magnetic moment,
 The values in the parenthesis are available measured data, collected from references 
 \cite{dasgupta2007low,raju1999transition,gardner2001neutron,gardner2010magnetic,
       poole2007magnetic,hodges2002first}.
 }
\label{magnetic}
\begin{ruledtabular}
\begin{tabular}{ccccccc}
     & L & S   & J    &  $\mu_S$   &  $\mu_L$ & $\mu_{tot}$   \\
\hline
\Eu  & 3 & 3   & 0    & 6.1 (---)~~& -1.1 (---)~~& 4.9 (0.6)  \\
\Gd  & 0 & 7/2 & 7/2  & 7.0 (---)~~& ~0.0 (---)~~& 7.0 (7.7)  \\
\Tb  & 3 & 3   & 6    & 6.0 (3.3)  & ~1.2 (1.7)  & 7.2 (5.1)  \\
\Dy  & 5 & 5/2 & 15/2 & 5.0 (4.9)  & ~3.3 (4.9)  & 8.2 (10.0) \\
\Ho  & 6 & 2   & 8    & 3.8 (3.9)  & ~---~(5.8)  & ---~(10.0) \\
\Er  & 6 & 3/2 & 15/2 & 3.0 (1.2)  & ~5.9 (1.9)  & 8.8 (3.2)  \\
\Yb  & 3 & 1/2 & 7/2  & 0.9 (0.3)  & ~1.5 (0.8)  & 2.4 (1.1)  \\
\end{tabular}
\end{ruledtabular}
\end{table*}

For more accurate comparison of the experimental and computational 
values of magnetic moments, we used the crystal electric field theory 
to decompose experimental moments into spin and orbital contributions.
Applying crystal electric field theory to the experimental neutron spectra,
the magnetic ground-state wave function of \A~pyrochlore compounds, except \Gd~and \Eu
are determined in terms of $|j, m_j\rangle$ basis set \cite{bertin2012crystal},
where $j$ and $m_j$ are total angular momentum and magnetic quantum number, respectively.
We used Clebsch-Gordan coefficients to decompose 
$|j,m_j\rangle$ to $|m_l,m_s\rangle$ states,
where $m_l$ and $m_s$ are orbital and spin magnetic quantum numbers, respectively.
The resulting magnetic ground-state wave functions in terms of $|m_l,m_s\rangle$ states
are given in appendix \ref{crystal-field},
and the obtained orbital and spin moments are presented in Table \ref{magnetic}.
The observed consistency between crystal field spin and orbital moments
and measured total moments confirms the obtained magnetic ground state wave functions.
It is seen that the calculated orbital moment of \Tb~and \Dy~are 
about 30\% smaller than the crystal field orbital moment,
while the calculated orbital moment of \Er~and \Yb
are significantly higher than the crystal field values.
On the other hand, the calculated spin moments show generally
more agreement with the corresponding crystal field values.
The obtained spin moments of \Dy~and \Ho~are very close to 
the crystal field data while in the case of \Tb, \Er, and \Yb
significant difference is visible between first-principles 
and crystal field spin moments.

\section{Conclusions}

The electronic structure and magnetic properties of \A~magnetic pyrochlores
were investigated by using full-potential density functional calculations 
within GGA+U scheme and relativistic spin orbit coupling (SOC).
The calculated band gaps are in good agreement with the available measured values.
We used \Tb~compound to show the possible multi-minima solutions
of the electronic structure of these magnetic pyrochlores.
It was argued that the half filling of the 4f shell of Gd significantly
decreases the contribution of f electrons in the valence shell of \Gd
and hence facilitates the electronic structure calculation of this system
to converge to its global minima. 
The calculated spin, orbital, and total magnetic moments of the systems 
within GGA+U+SOC calculations were presented and compared with the available
measured total magnetic moments.
The best agreement was observed in \Gd, while \Eu~and \Er
compounds showed the highest deviation from experiment.
The phenomenological magnetic ground state wave function 
of \Tb, \Dy, \Ho, \Er, and \Yb~compounds were determined in 
the framework of crystal electric field theory for decomposition of
the measured total magnetic moments into spin and orbital contributions. 
Comparing the calculated and the crystal field spin and orbital moments 
shows more accuracy of the calculated spin moments.

\section*{Acknowledgments}

This work was jointly supported by the Vice Chancellor
for Research Affairs of Isfahan University of Technology (IUT),
and ICTP Affiliated Centre at IUT.
M. A. is grateful to Farhad Shahbazi for helpful discussions.

\appendix

\section{Crystal field analysis}
\label{crystal-field}

The magnetic ground-state wave function of \Aa~pyrochlore compounds in 
terms of $|j,m_j\rangle$ states are as follows~\cite{bertin2012crystal}:

\begin{eqnarray}
|\phi_g^{Tb}\rangle=
+0.96|\pm4\rangle\pm0.13|\pm1\rangle \non \\
-0.12|\mp2\rangle\mp0.23|\mp5\rangle \non 
\end{eqnarray}
\vspace{-0.5cm}
\begin{eqnarray}
|\phi_g^{Dy}\rangle=
+0.981|\pm\frac{15}{2}\rangle\pm0.190|\pm\frac{9}{2}\rangle \non \\
+0.022|\pm\frac{3}{2}\rangle\pm0.037|\mp\frac{3}{2}\rangle  \non \\
+0.005|\mp\frac{9}{2}\rangle\pm0.001|\mp\frac{15}{2}\rangle        \non
\end{eqnarray}
\vspace{-0.5cm}
\begin{eqnarray}
|\phi_g^{Ho}\rangle=
-0.979|\pm8\rangle\pm0.189|\pm5\rangle \non \\
-0.014|\pm2\rangle\pm0.070|\mp1\rangle \non \\
-0.031|\mp4\rangle\pm0.005|\mp7\rangle \non
\end{eqnarray}
\vspace{-0.5cm}
\begin{align}
|\phi_g^{Yb}\rangle=&
+0.376|\pm\frac{7}{2}\rangle+0.922|\pm\frac{1}{2}\rangle \non \\
&-0.093|\mp\frac{5}{2}\rangle  \non
\end{align}
\begin{align}
|\phi_g^{Er}\rangle=&
+0.47|\pm\frac{13}{2}\rangle\pm0.42|\pm\frac{7}{2}\rangle  \non \\
 &-0.57|\pm\frac{1}{2}\rangle\mp0.24|\mp\frac{5}{2}\rangle  \non \\
 &+0.47|\mp\frac{11}{2}\rangle \non
\end{align}
\vspace{-0.5cm}

We used Clebsch-Gordan coefficients to represent these magnetic ground-state
wave functions in terms of $|m_l,m_s\rangle$ states:

\begin{align}
|\phi_g^{Tb}\rangle=&
+0.46|-3,-1\rangle + 0.71|-2,-2\rangle\non \\
&+0.46|-1,-3\rangle - 0.04|+1,-2\rangle\non \\
&-0.08|~0,-1\rangle - 0.08|-1,~0\rangle\non \\
&-0.04|-2,+1\rangle - 0.01|+2,-3\rangle\non \\
&-0.01|-3,~2\rangle  - 0.02|~3,-1\rangle\non \\
&-0.06|~2,~0\rangle  -0.08|~1,~1\rangle\non\\ 
&-0.06|~0,~2\rangle  -0.02|-1,~3\rangle\non \\
&+0.16|~3,~2\rangle\non+0.16|~2,~3\rangle \non
\end{align}

\begin{align}
|\phi_g^{Dy}\rangle=&
+0.9800|\pm5,\pm\frac{5}{2}\rangle\non 
\pm0.0970|\pm2,\pm\frac{5}{2}\rangle\non \\
&\pm0.1330|\pm3,\pm\frac{3}{2}\rangle\non 
\pm0.0890|\pm4,\pm\frac{1}{2}\rangle\non \\
&\pm0.0280|\pm5,\mp\frac{1}{2}\rangle\non
\mp0.0010|\mp4,\pm\frac{5}{2}\rangle\non \\
&\pm0.0080|\mp3,\pm\frac{3}{2}\rangle\non 
\pm0.0180|\mp2,\pm\frac{1}{2}\rangle\non \\
&\pm0.0240|\mp1,\mp\frac{1}{2}\rangle\non
\pm0.0180|~0,\mp\frac{3}{2}\rangle\non \\
&\pm0.0070|\pm1,\mp\frac{5}{2}\rangle\non 
-0.0045|\mp1,\pm\frac{5}{2}\rangle\non \\
&-0.0110|~0,\pm\frac{3}{2}\rangle\non
-0.0142|\pm1,\pm\frac{1}{2}\rangle\non \\
&-0.0107|\pm2,\mp\frac{1}{2}\rangle\non 
-0.0046|\pm3,\mp\frac{3}{2}\rangle\non \\
&-0.0098|\pm4,\mp\frac{5}{2}\rangle\non
+0.0007|\mp5,\pm\frac{1}{2}\rangle\non \\
&+0.0023|\mp4,\mp\frac{1}{2}\rangle\non
+0.0035|\mp3,\mp\frac{3}{2}\rangle\non \\
&+0.0025|\mp2,\mp\frac{5}{2}\rangle\non
\pm0.0010|\mp5,\mp\frac{5}{2}\rangle\non
\end{align}

\begin{align}
|\phi_g^{Ho}\rangle=&
-0.9790|\pm6,\pm2\rangle \pm 0.1184|\pm3,\pm2\rangle\non \\
&\pm0.1290|\pm4,\pm1\rangle \pm 0.0670|\pm5,~0\rangle\non \\
&\pm0.0160|\pm6,\mp1\rangle - 0.0047|~0,\pm2\rangle\non \\
&-0.0088|\pm1,\pm1\rangle - 0.0080|\pm2,~0\rangle\non\\
&-0.0046|\pm3,\mp1\rangle - 0.0012|\pm4,\mp2\rangle\non \\
&\pm0.0098|\mp3,\pm2\rangle \pm 0.0300|\mp2,\pm1\rangle\non\\
&\pm0.0450|\mp1,~0\rangle \pm 0.0400|~0,\mp1\rangle\non \\
&\pm0.0180|\pm1,\mp2\rangle - 0.0007|\mp6,\pm2\rangle\non\\
&-0.0050|\mp5,\pm1\rangle - 0.0210|\mp3,\mp1\rangle\non \\
&-0.0160|\mp2,\mp2\rangle - 0.0140|\mp4,~0\rangle\non\\
&\pm0.0025|\mp6,\mp1\rangle \pm 0.0043|\mp5,\mp2\rangle\non
\end{align}

\begin{align}
|\phi_g^{Yb}\rangle=&
+0.37|\pm3,\pm\frac{1}{2}\rangle\non\pm0.70|~0,\pm\frac{1}{2}\rangle\non \\ 
&\pm0.60|\pm1,\mp\frac{1}{2}\rangle\non-0.03|\mp3,\pm\frac{1}{2}\rangle\non \\
&-0.09|\mp2,\mp\frac{1}{2}\rangle\non
\end{align} 

\begin{align}
|\phi_g^{Er}\rangle=&+0.21|\pm6,\pm\frac{1}{2}\rangle\non 
+0.42|\pm5,\pm\frac{3}{2}\rangle    \non \\
&\pm0.04|\pm5,\mp\frac{3}{2}\rangle \non 
\pm0.16|\pm4,\mp\frac{1}{2}\rangle  \non \\
&\pm0.31|\pm3,\pm\frac{1}{2}\rangle \non 
\pm0.25|\pm2,\pm\frac{3}{2}\rangle  \non \\
&-0.16|\pm2,\mp\frac{3}{2}\rangle   \non 
-0.35|\pm1,\mp\frac{1}{2}\rangle    \non \\
&-0.38|~0,\pm\frac{1}{2}\rangle     \non
-0.20|\mp1,\frac{3}{2}\rangle       \non \\
&\mp0.12|\mp1,\mp\frac{3}{2}\rangle \non
\mp0.17|\mp2,\mp\frac{1}{2}\rangle  \non \\
&\mp0.11|\mp3,\pm\frac{1}{2}\rangle \non
\mp0.04|\mp4,\pm\frac{3}{2}\rangle  \non \\
&+0.37|\mp4,\mp\frac{3}{2}\rangle   \non 
+0.28|\mp5,\mp\frac{1}{2}\rangle    \non \\
&+0.08|\mp6,\pm\frac{1}{2}\rangle   \non
\end{align}

For calculating the spin and orbital magnetic moments from the above 
wave functions, one should take into account that the \Tb, \Dy, \Ho, and \Yb
are Ising like systems while \Er~adopts $\mathrm{XY}$ model magnetism. 
For Ising like systems, $g_l\langle L_z\rangle$ and $g_s\langle S_z\rangle$ 
are calculated as orbital and spin moments of the magnetic atom of the systems,
while in $\mathrm{XY}$-like systems, $\langle S_\pm\rangle \gg \langle S_z\rangle$ and  
$\langle L_\pm\rangle \gg \langle L_z\rangle$, 
therefore $g_l\langle L_\pm\rangle$ ($L_\pm=L_{x}\pm{iL_y}$,$g_l=1$) 
and $g_s\langle S_\pm\rangle$ ($S_\pm=L_{x}\pm{iL_y}$,$g_s=2$)
are calculated as orbital and spin moments, respectively.

\bibliography{bibliography}
\end{document}